\documentclass[aip,jap,reprint]{revtex4-1}
\usepackage{graphicx,float,amssymb,subfigure,dcolumn,epstopdf}
\usepackage{multirow}
\usepackage{bm}
\begin{document}
	 \title{Band gap engineering of PtSe$_2$  }
	 \author{ Zhisuo Huang}
	 \affiliation{State Key Laboratory of Electronic Thin Films and Integrated Devices,
	 	University of Electronic Science and Technology of China, Chengdu, 610054, P. R. China}

	 \author{Wenxu Zhang\footnote{Corresponding author. E-mail address: xwzhang@uestc.edu.cn}}
	 \affiliation{State Key Laboratory of Electronic Thin Films and Integrated Devices,
	 	University of Electronic Science and Technology of China, Chengdu, 610054, P. R. China}
	 \author{ Wanli Zhang}
	 \affiliation{State Key Laboratory of Electronic Thin Films and Integrated Devices,
	 	University of Electronic Science and Technology of China, Chengdu, 610054, P. R. China}
	 \date{\today}
	 \begin{abstract}
	 	Besides its predicted promising high electron mobilities at room temperature, PtSe$_2$ bandgap sensitively depends on the number of monolayers combined by van der Waals interaction according to our calculations. We understand this by using bandstructure calculations based on the density functional theory. It was found that the front orbitals of VBM and CBM are contributed mainly from $p_z$ and $p_{x+y}$ orbitals of Se which are sensitive to the out-plane and in-plane lattice constants, respectively. The van der Waals force enhances the bonding out-of-plane, which in-turn influences the bonding in-plane. We found that the thickness dependent bandgap has the same origin as the strain dependent bandgap, which is from the change of the front orbital interactions. The work shows the flexibilities of tuning the electronic and optical properties of this compound in a wide range.
	 \end{abstract}
	 \maketitle
	 \section{introduction}
	Bandgap engineering is an important and interesting aspect in semiconductor consortium. Being a fundamental property of semiconductors, it determines all the applications of the materials including electron transport, optical absorption, thermal stability, etc. In newly discovered 2D  semiconductors, graphene has received a great deal of researches to demonstrate its applications. One main obstacle to use it in logical devices is to open the bandgap. Quite some strategies have been demonstrated, like applying in-plane strain \cite{C5RA03422A}, choice of suitable substrates, such as hexagonal boron nitride \cite{PhysRevB.76.073103}, and selective absorption of atomic hydrogen \cite{balog}. However, the effects are usually very small.
	Tuning the bandgap of MX$_2$ type transitional metal dichalcogenide (TMD)  materials shows more successful stories both in the theoretical and experimental aspects, where pressures \cite{PhysRevB.85.033305, PhysRevLett.113.036802, PhysRevB.87.125415}, van der Waals interactions \cite{PhysRevB.85.033305, PhysRevB.83.245213, PhysRevB.88.075421, PhysRevB.84.045409} etc. are shown to be an effective way to engineer the bandgap. The change of the electronic properties with the layer numbers was proposed to be the effect of quantum confinement \cite{PhysRevB.83.245213}, the structure variations \cite{PhysRevB.88.075421} and spontaneous electrical polarization at the interface of the heterostructures \cite{Kou2013}. Systematic studies of bandgap evolutions with the layer number in group-VIB TMDs by Zunger \textit{et al.}\cite{Zhang2015} reveals that the indirect-to-direct band gap transformation is triggered not only by (kinetic-energy controlled) quantum confinement but also by (potential-energy controlled) band repulsion and localization. It was shown  \cite{PhysRevB.86.075454} that the band gap of bilayer sheets of semiconducting TMDs can be reduced smoothly by applying vertical compressive pressure. The semiconductor-metal (S-M) transition is attributed to the lifting of the degeneracy of the bands at the Fermi level caused by interlayer interactions via charge transfer from metallic atoms to chalcogens. Bradley \textit{et al.} \cite{Bradley2015} found that the electronic (quasiparticle) bandgap can decrease nearly 1.0 eV when going from one layer to three due to interlayer coupling and screening effects, which change the electronic wave function hybridization in the valleys of both the valence and conduction bands.
	\par  We  predicted that PtSe$_2$ may have high electron mobility at room temperature\cite{Zhang2014}, if only long wave length acoustical phonon scattering is included. It may have the highest phonon limited electron mobility when more scattering sources are considered\cite{1505.05698}. Recently, monolayer PtSe$_2$ film was synthesized \cite{Wang2015} and its photocatalytic activity is evaluated by a methylene-blue photodegradation experiment, demonstrating its practical application as a promising photocatalyst. Moreover, circular polarization calculations predict that monolayer PtSe$_2$ has also potential applications in valleytronics\cite{Wang2015}. Regarding to these interesting application of the material, the bandgap is very crucial. In this work, we demonstrate that the bandgap, as well as the effective masses of PtSe$_2$ can be largely tuned by the number of layers. The reason of the changes can be due to the interlayer interactions via van der Waals (vdW) forces when the distances of them are changed. The van der Waals forces increase with the number of layers are the main causes of this variation. There is no electrons between the layers, however, the vdW exists because they are predicted on induced dipole interactions rather than static charge interactions.


    \section{Calculation details}
    The calculations were performed mainly with pseudo-potential (PP) code PWscf\cite{PWscf}, with the vdW-DFT module to treat van der Waals interaction between layers\cite{vdW-DFT,vdW-DFT-2,vdW-DFT-3}. The LDA functional was chosen to be that parameterized by Perdew and Zunger (PZ)\cite{PP-PZ}, and the GGA functional parameterized by Perdew, Burke,  and Ernzerhof (PBE)\cite{PP-PBE-1,PP-PBE-2,PP-PBE-3}. The plane-wave kinetic energy cutoff was set to 70 Ry with the density cut-off of 700 Ry, and  shifted $13\times13\times1$ Monkhorst-Pack meshes for the layered structures and $13\times13\times8$ for the bulk were used to perform Brillouin zone integration in order to ensure the convergence of the results. The convergence of the total energy was set to be better than  10$^{-9}$ Hartree. A vacuum layer with thickness of 50 a.u. was used to model the 2D-nature of the compounds. Forces on the atoms are limited within 0.001 Ry/a.u. after full geometry relaxation. Crosscheck of the  LDA and GGA calculations was performed by the Full Potential Local Orbital codes (FPLO) \cite{fplo} with the accuracy control set at the same level.

\section{Results and discussions}
\subsection{Geometric trends with different functionals}
    Lattice information of bulk PtSe$_2$ are calculated with full geometry relaxation as shown in Table. \ref{table:crystal-bulk}, compared with experiments and theoretical calculations. The in-plane lattice constant $a$ agrees well between the experiment and calculations as can be seen from the Table. All the functionals reproduce the experiments with accuracy within 3\%. The distance between Se and Pt within the unit cell of PtSe$_2$  $d_{\text {Se-Pt}}$ is extremely accurate reproduced by LDA, where the discrepancy is more than 3\%  within PBE and PBE+vdW.
    It is well known that PBE does not describe the dispersive van der Waals interactions, which gives more than 26\% overestimation of the  distance $c$ between  PtSe$_2$ monolayer as in our work and other similar works, like that of Piotrowski\cite{PhysRevB.88.075421}, while PBE+vdW can  correct this value to 16.6\%. It is notable that PBE+D3, another dispersive functional in the  '{\em Jacob's ladder}' \cite{Jacobs}, gives over-binding about -6.5\%, which is the same as LDA. Regarding to these results, there is still much room to improve the accuracy of including proper dispersive functionals in these van der Waals crystals.

\begin{table}
	\centering
	\caption{Crystal parameters under LDA, PBE and PBE+vdW. $a$ is the in-plane lattice constant, $c$ is the distance of Pt atoms in $z$-direction, and $d_{\text{Se-Pt}}$ is the distance between the neighboring Se and Pt planes.}\label{table:crystal-bulk}
	\begin{tabular}{lllc}
		\hline
		 & $a$ (\AA)    & $c$ (\AA)     & $d_{Se-Pt} $ (\AA) \\
		\hline
		Exp. \cite{PtX2,PtX2-2,Struct-sole}    & 3.728 & 5.081 & 1.270 \\
		LDA     & 3.757 & 4.759& 1.271 \\
		PBE    & 3.764, 3.75 \cite{PhysRevB.88.075421}  & 6.433, 6.55 \cite{PhysRevB.88.075421} & 1.314  \\
		PBE+vdW & 3.814, 3.79\cite{PhysRevB.88.075421}  & 5.927,4.75 \cite{PhysRevB.88.075421} & 1.356 \\
		\hline
	\end{tabular}
\end{table}

\begin{figure}
	\includegraphics[scale=0.25]{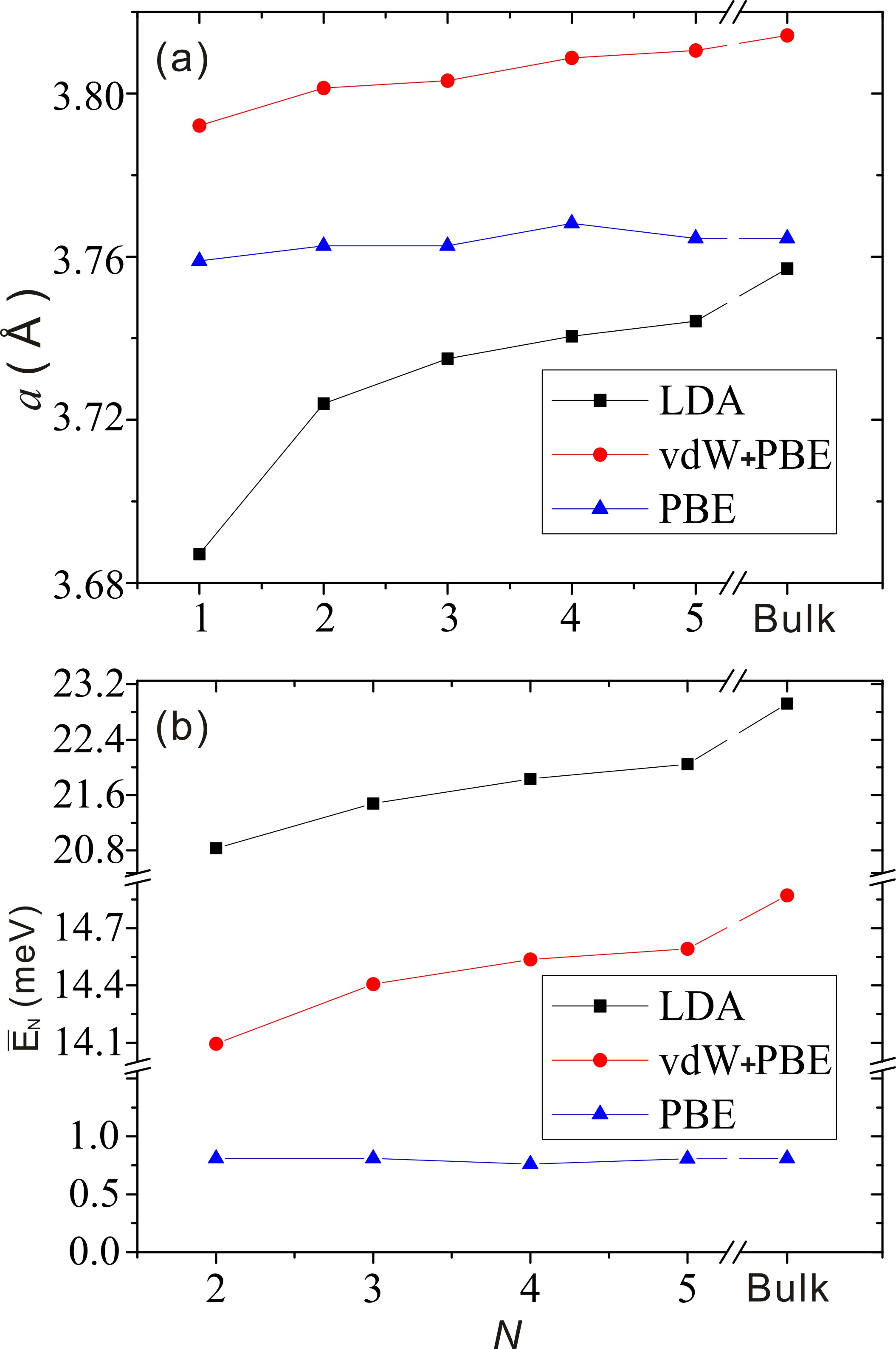}
	\caption{\label{fig:Vary_a-c-Ei_layer} Trends of lattice constant $a$ with layer number $N$ varying from 1 to 5 and bulk (a) and the averaged binding energy $\bar{E}_N$ (b) .}
\end{figure}
\par  The in-plane lattice constant $a$ increases with the additional layers, and shoots gradually to the bulk value (Fig.\ref{fig:Vary_a-c-Ei_layer} (a)). Among the different functionals, the variations of LDA are the most significantly, while values from PBE and PBE+vdW show only marginal variations. The LDA gives the smallest in-plane values because its intrinsic overbinding characters. Compared with the bare PBE, the PBE+vdW gives larger in-plane lattice because it corrects the out-of-plane lattice constant to a smaller values.
The averaged inter-layer interaction energy $\bar{E}_N$ is calculated by Equ. (\ref{equ:average-interaction-energy}),
\begin{equation}
\bar {E}_N=\frac{NE_{1}-E_N}{N-1}, \label{equ:average-interaction-energy}
\end{equation}
where the $N (\ge 2)$ is the number of layers, and $E_N$ is the total energy of $N$-layers. The variation is shown in Fig.\ref{fig:Vary_a-c-Ei_layer} (b). The PBE gives the lowest value, and LDA gives the largest value. The bonding energy correction by vdW to bare PBE is about 13 meV, while LDA overestimates the value by 50\%. At the same time, we can see that the averaged bonding energy increases gradually with the increase of the layer number.
%
\subsection{Electronic structure variations with the number of layers }

\begin{figure}
	\includegraphics[angle=90,width=0.45\textwidth]{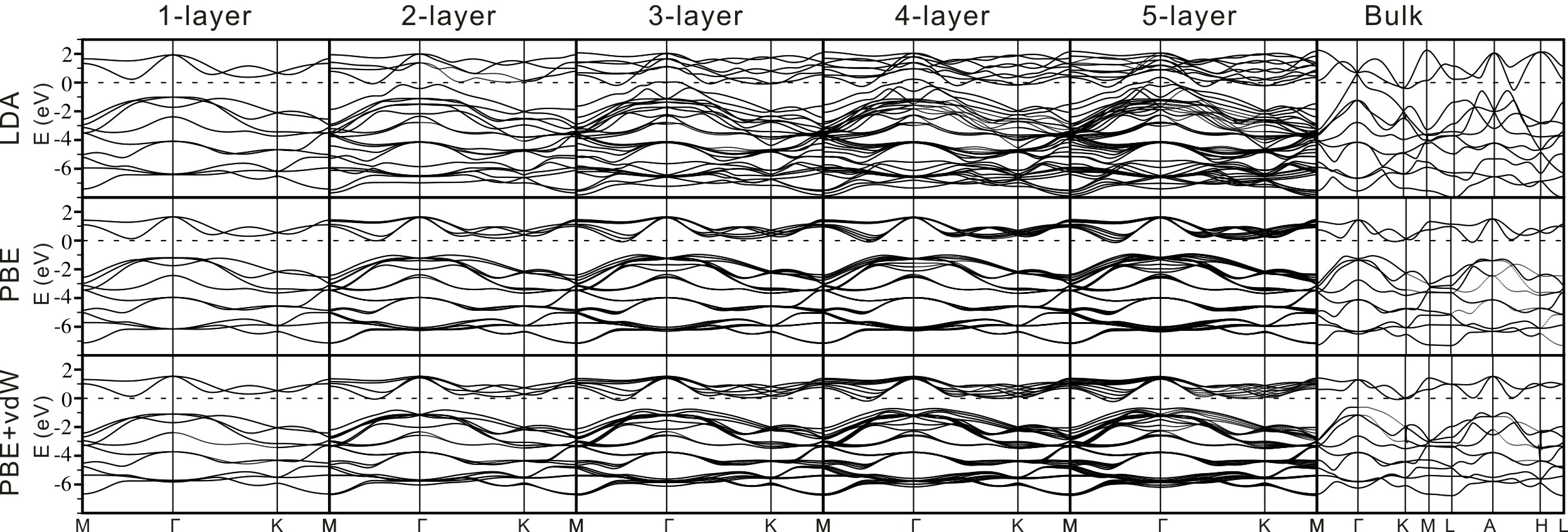}
	\caption{\label{fig:Band_alignment} Band alignment using different layers under different xc-functionals. The Fermi level is shown by the dashed lines and set the 0 eV.}
\end{figure}

\par It is nontrivial to compare the band structure of the compounds with different thickness because it requires the alignment of the band structures at different $N$ with respect to each other. In this work, the band alignment is obtained with respect to the work function of different number of layers. The aligned band structures are shown in Fig.\ref{fig:Band_alignment}. Experimentally, bulk PtSe$_2$ is a semi-metal as measured from X-ray and densitometric techniques \cite{PtX2-2,Struct-sole} which was confirmed by far-infrared (4000 - 40 cm$^{-1}$) and X-ray techniques \cite{Struct-kliche}. According to the DFT calculations, the bandgap is dependent on the xc-functionals. As can be seen in the figure, bulk PtSe$_2$ is metal under LDA, while it has a finite indirect bandgap with PBE and PBE+vdW. Guo and Liang\cite{PtX2} calculated platinum dichalcogenide band structures and showed that bulk PtSe$_2$ is semi-metal, consistent with the results by Dai \emph{et al. } using GGA\cite{bulk-PtSe2}. But these calculations were performed with the experimental crystal structure. On the other hand, Piotrowski \emph{et al.}\cite{PhysRevB.88.075421} used fully relaxed lattice of PtSe$_2$ and show that PtSe$_2$ under PBE is semiconductor, which is consist with our PBE result, while it is metal with dispersion corrections (PBE+D3), corresponding to the result of LDA. However, under PBE+vdW bulk PtSe$_2$ is a semiconductor with smaller bandgap than that of PBE, which indicating that the vdW-DFT correction is not enough for fixing the van der Walls forces. As will be argued later in this work, the differences lie in the different lattice constant under different functionals, while the bandgap is a sensitive to the variation of the bond length.

\begin{figure}
	\includegraphics[scale=0.7]{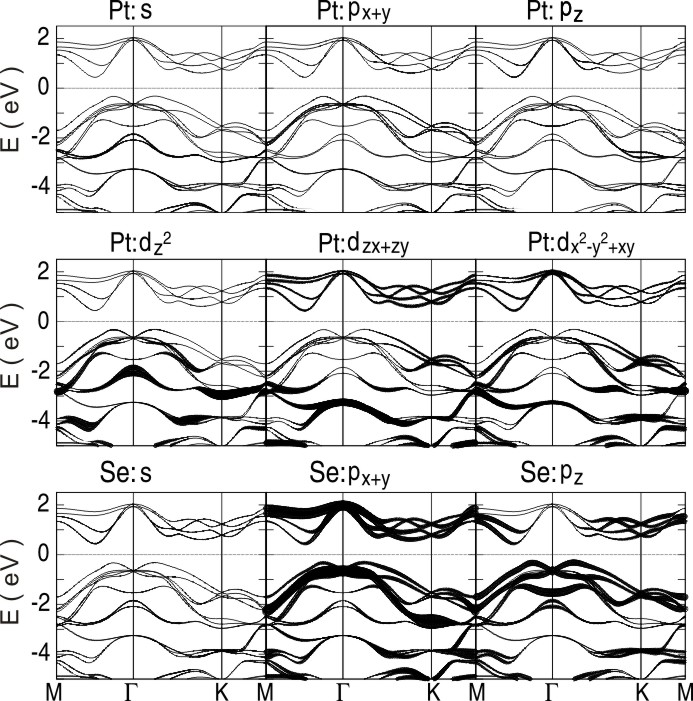}
	\caption{\label{fig:fat_band} Fat band representations for the s, p, d orbital of Pt and s, p orbital of Se under PBE+vdW correction of double layered PtSe$_2$.}
\end{figure}

\par In order to get a better insight into the contributions to the states by different electron orbitals, we plot the fat-bands of double layers as shown in Fig.\ref{fig:fat_band} as an example. The bulk band structures are well analyzed by Dai \cite{bulk-PtSe2}. The bandgap of the bulk is developed between the $K$ and $\Gamma$ points \cite{bulk-PtSe2}. It is shown that the main contributions of the state of the valence band maximum (VBM) at $\Gamma$ are from Se $p_z$ orbitals, while the conduction band minimium (CBM) at K points is mainly contributed by $p_{x+y}$ states from Se as well as $d_{xz+yz}$ states from Pt. The consequence of this electronic configuration is that the bandgap is sensitive to both in-plane and interplane lattice constant. However, the CBM and VBM are different from the bulk in the finite layer 2D crystals, where the VBM is located in the middle of $\Gamma-M$ and VBM is located around the $\Gamma$ point, which simulates the dispersions the path of $L-A-H-L$  with $k_z=\frac{1}{2}$ in the bulk.
\begin{figure}
	\includegraphics[width=0.35\textwidth]{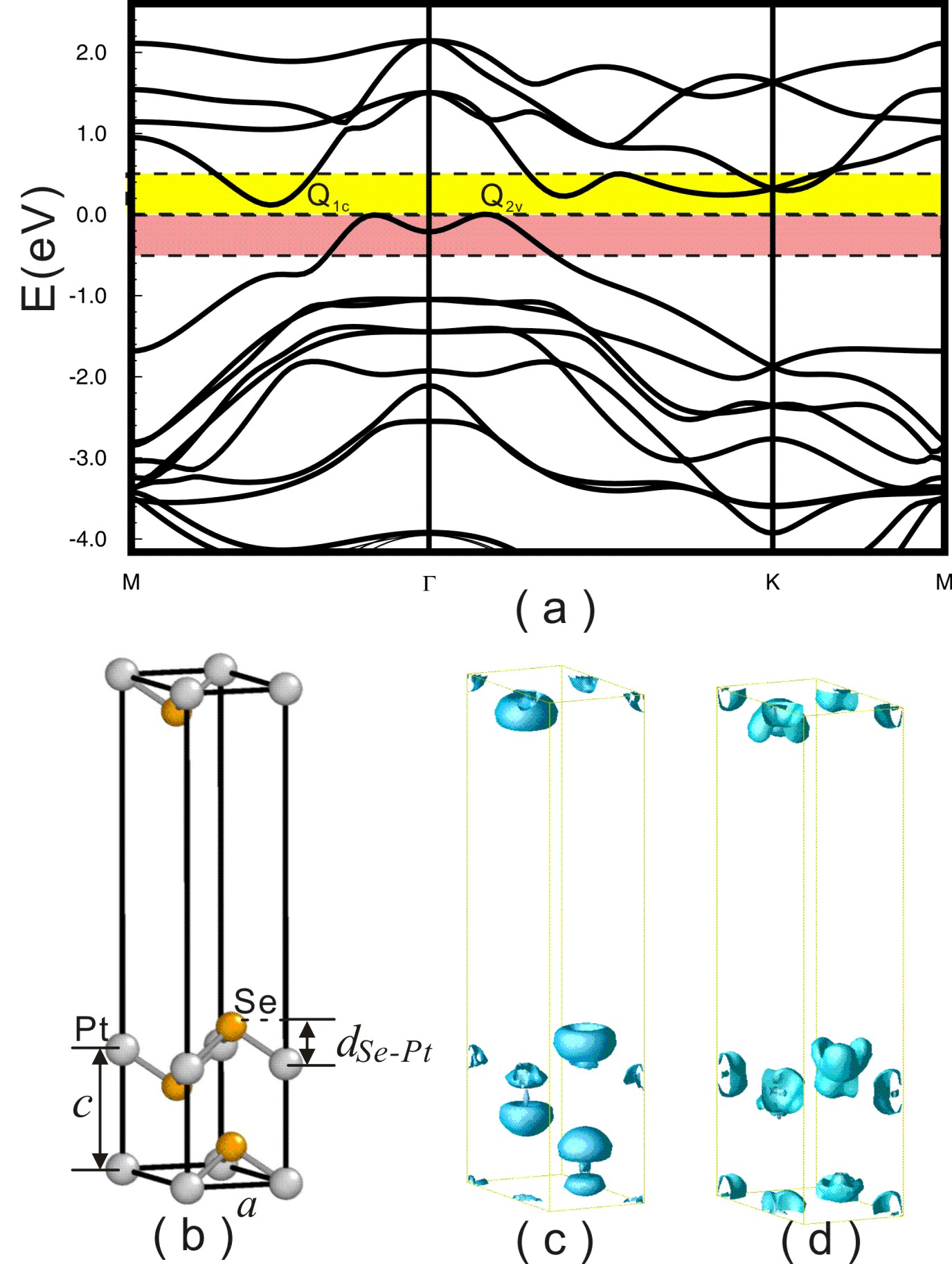}
	\caption{\label{fig:iso-2l} (color online) The LDA band structure of 2 layer PtSe$_2$(a), with the Fermi level set at $E=0$ eV. The unit cell used in the calculation (b). The electron density isosurface of the front states of VBM (c) and CBM (d) within the energy window of 0.5 eV from the Fermi level as shown in (a). }
\end{figure}

\par The electron density isosurface of the front orbitals within energy range of 0.5 eV, i.e. within the dashed lines in Fig.\ref{fig:iso-2l}(a) from the CBM and VBM of double layers are plotted in Fig.\ref{fig:iso-2l}(c) and (d). The positions of the corresponding atoms are shown in Fig.\ref{fig:iso-2l}(b). The electronic states of VBM mimic the electron distributions of $p_z$ orbitals which show the main lobe in the $z$ direction of Se. The CBM isosurfaces show the in-plane anisotropy clearly. The one from Se points to Pt, indicating interactions between these orbitals within layer. When more layers are added, as seen before in Fig.\ref{fig:Vary_a-c-Ei_layer} (a), the van der Waals forces introduce compression in $c$ and increase in $a$. This will induce stronger interaction of the bonding of $p_z$ orbitals and less bonding between $p_{x+y}$ orbitals. Thus the energy of antibonding state are lowered and the bonding $p_z$ orbitals are increased. This is the reason for the reduction of the bandgap. The electrons around $\Gamma$ are from the antibonding state of $p_z$ orbitals. Therefore, decrease of $z$ will increase its eigenenergy. Meanwhile, the states around $K$ and $Q_{1c}$ are antibonding of $p_{x+y}$ orbitals, therefore, increase of the lattice decreases the energy. This is the scenario for bandgap variations with different number of layers and strains.


\par
\begin{figure}
\includegraphics[scale=0.25]{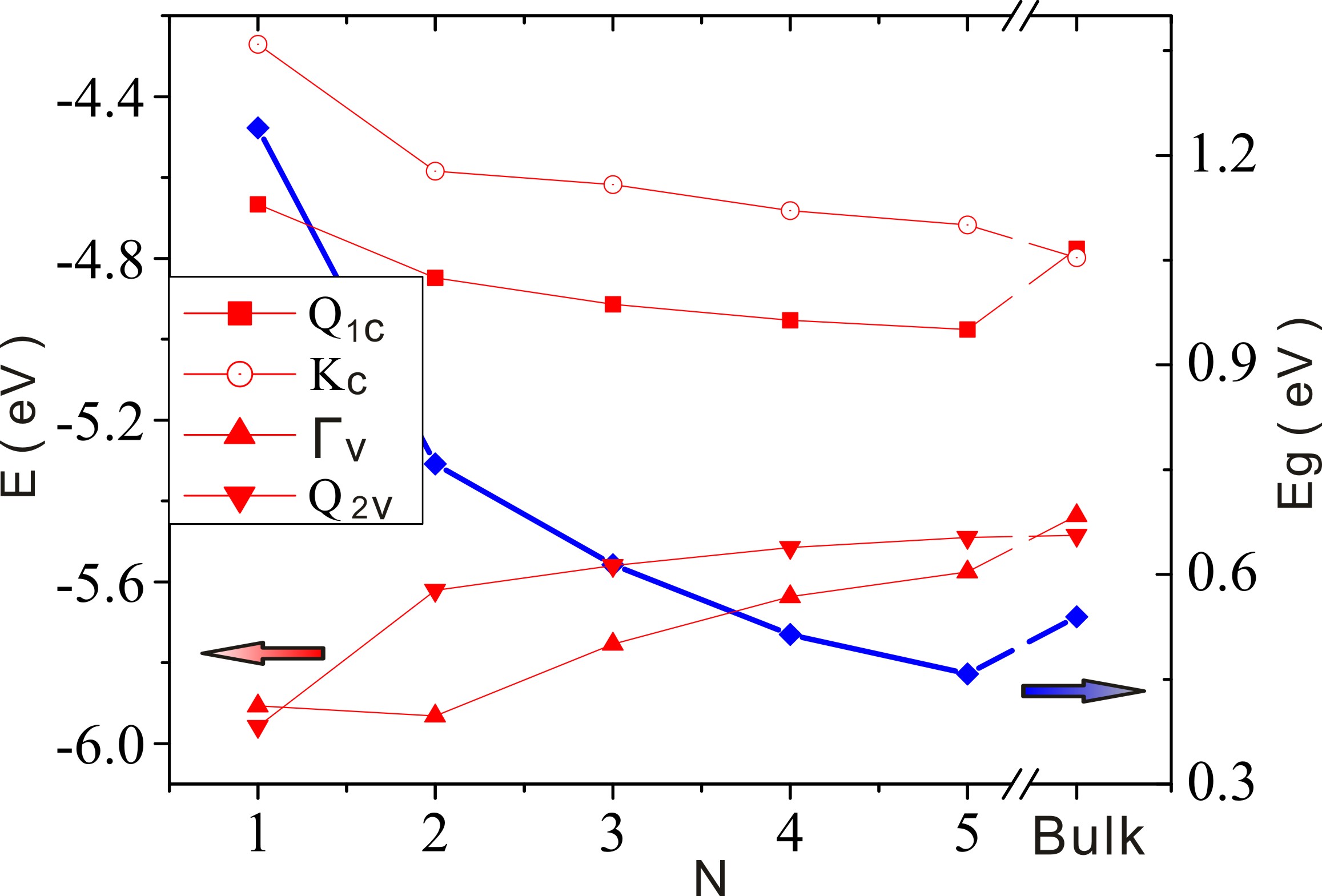}
\caption{\label{fig:E_level-Eg_abs} (color online) Trend of $Q_{1c}$, $Q_{2v}$(as shown in Fig. \ref{fig:iso-2l} (a)), $K_{c}$, $\Gamma_{v}$ and band gap with different layers under vdw-DFT (solid blue line).}
\end{figure}

 The variation of the eigenvalues at the different k-points with respect to the number of layers, as well as the band gaps,  are in Fig.\ref{fig:E_level-Eg_abs}. As we can see, the VBM is located at the $Q_{2v}$, between M and $\Gamma$ in 2-5 layered materials, while in monolayer and bulk it is at the $\Gamma$ as also shown in Fig.\ref{fig:Band_alignment} in PBE+vdW calculations. As for the CBM, it is the same in all the cases. Only in bulk, the eigenvalue at K-point is degenerated with $Q_{1c}$ due to band repulsion. Within the cases we considered, there is no semiconductor-metal transitions, however the bandgap shrinks from 1.2 eV to 0.5 eV when the layer number is increased from monolayer to five-layered structure, indicating the tendency of S-M transitions.
\par
\begin{figure}
\includegraphics[scale=1.8]{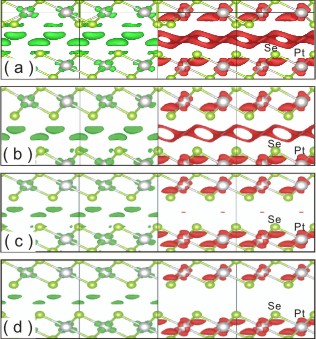}
\caption{\label{fig:charge_all} (color online) Isosurfaces of charge accumulation (green) and depletion (red) of PtSe$_2$ with N layers, N=2(a), N=3(b), N=4(c) and N=5(d).}
\end{figure}

 The change of the eigenenergy at the special k-points is the results of the electron interactions with different number of layers. This can be observed in Fig.\ref{fig:charge_all} where the corresponding charge accumulation and depletion isosurfaces in layers with different $N$'s are depicted, from which two conclusions could be drawn: With the layer number increment, charges are becoming more localized no matter in intra-layers or in inter-layers, and the charges of inner Se atoms are transforming to Pt atoms, while that of outer Se atoms is hardly influenced. These charge behavior is similar with that in the study when consider strain-induced S-M transitions\cite{PhysRevB.86.075454, Strain-induced}.

\par The change of the electron distribution and consequently the change of the bandgap are stemmed from the relaxation of the interlayer distances when the number of the layers are changed. At this stage, a primary conclusion can be drawn that the van der Waals interaction between layers when layered-PtSe$_2$ stacking to bulk PtSe$_2$ induces a positive in-plane strain (in-plane crystal constant $a$ increases) and a negative out-plane strain (distance between layers decreases), which leads to the transfer of the electrons from $p_{z}$ orbitals to in-plane orbitals: $p_{x+y}$ and $d_{xz+yz}$. As a consequence, the CBM goes down and the VBM goes up as discussed above, narrowing the band gap.
\subsection{Closing the bandgap by in-plane strains }
 \par
\begin{figure}
\includegraphics[scale=0.5]{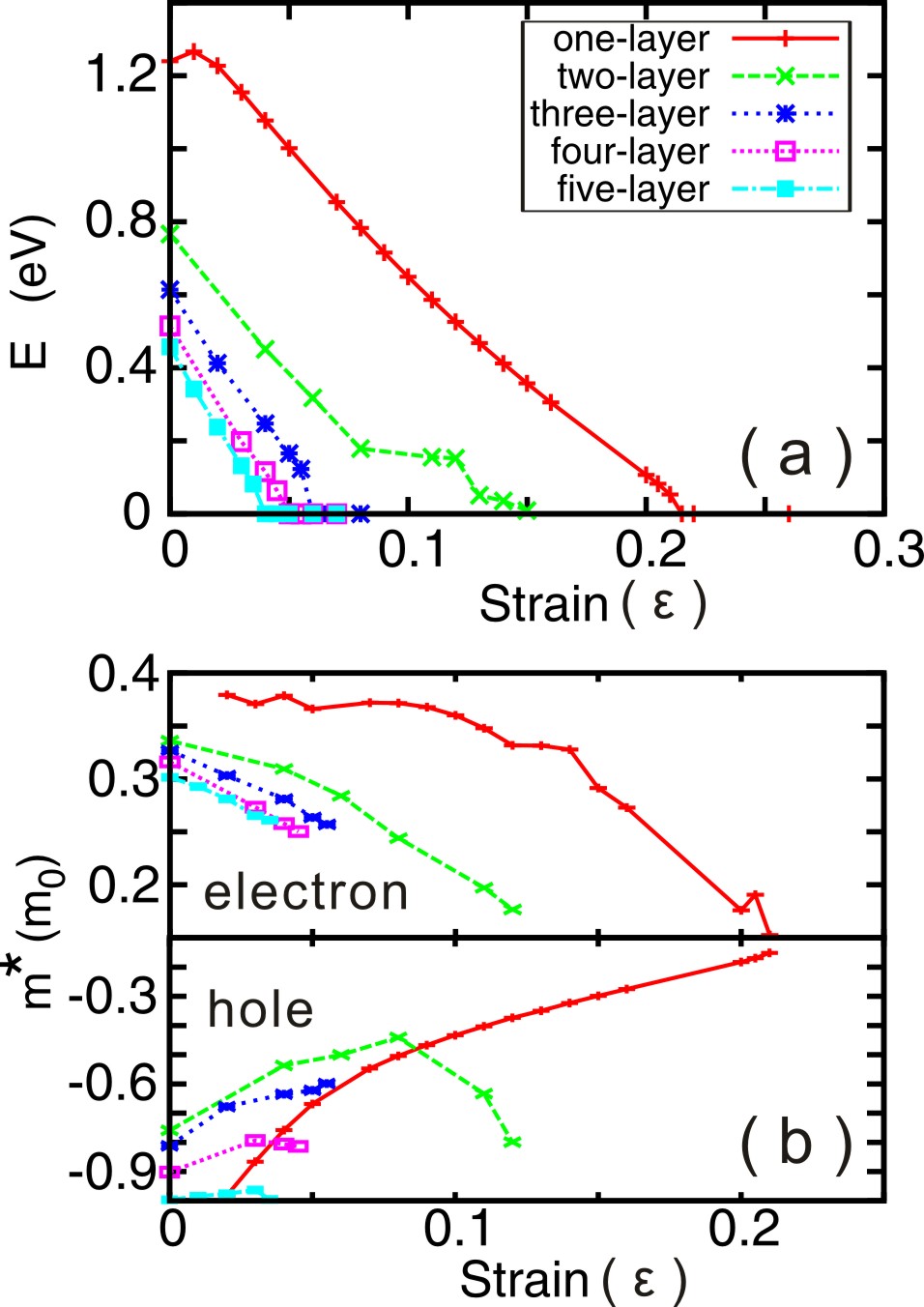}
\caption{\label{fig:Strain-bandgap} (color online) Bandgap engineering with in-plane strain ($\varepsilon$) under PBE+vdW correction in 1-5 layers (a), and effective mass of electron (upper) and hole (lower) (b) with the same logo of (a).}
\end{figure}

\begin{figure}
\includegraphics[scale=0.75]{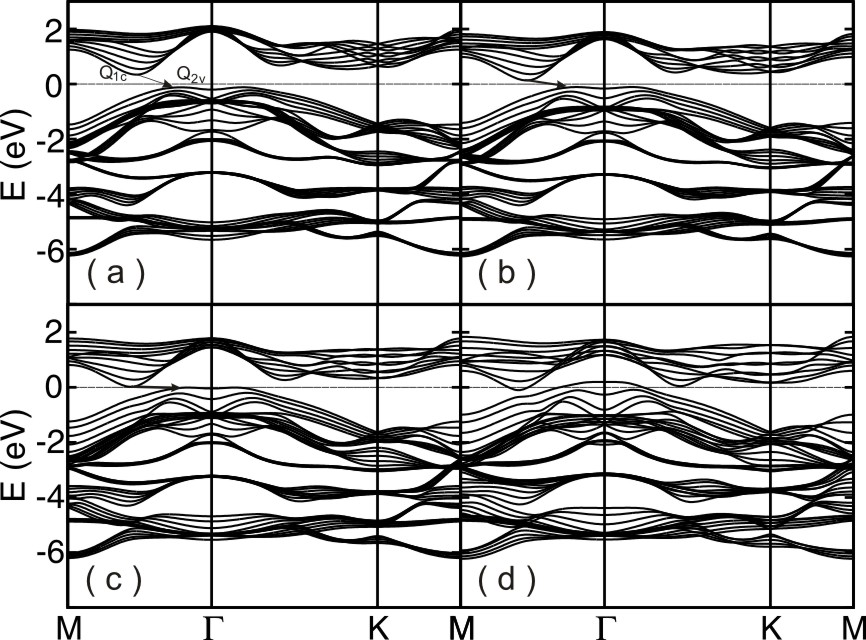}
\caption{\label{fig:Strain-bands-dos} Band structure with in-plane strains $\varepsilon=$ 0.00 (a), 0.02(b), 0.04(c) and 0.06(d) five-layered structure with PBE+vdW.}
\end{figure}

  Applying stains is an effective way to tune the carrier properties as well established in conventional semiconductors Si etc.. Here, in-plane homogeneous strains can be applied in these layered crystals to induce semiconducting-metallic transition. As shown in Fig. \ref {fig:Strain-bandgap} (a), expanding the in-plane lattice constant, which is equivalent to a pressure applied on the films, reduces the bandgap. Its electronic origin is the same as the layer number dependent bandgap: Positive in-plane strain weakens the interactions between Se and Pt. With the increase of the number of layers, the strain required to close the gap is reduced to about only 4\% when the number of layers is 5, of which the band structures under strains of 0.00, 0.02, 0.04 and 0.06 are shown in Fig.\ref{fig:Strain-bands-dos} in which VBM is changed from $Q_{v2}$ in the vincinity of $\Gamma$ to $\Gamma$. At the same time, both the effective mass of electrons and holes decrease with the increase of strains, except the holes in 2-layered compounds. Without strains, increase the number of layer will increase the hole $m^{*}$ and decrease the electron $m^{*}$, and monolayer is, however, an exception because the VBM is located at $\Gamma$.


\section{Conclusions}
In summary, PtSe$_2$ with the number of layers increasing from 1 to 5 was studied by PBE+vdW with fully relaxed atomic positions. The trends of the band structure and the atomic positions with the number of layers were analysed. By comparison of the band and charge redistribution of different layers, we traced out the role of van der Waals forces plays in these materials. When stacking layers together, the van der Waals induces a positive in-plane strain (in-plane crystal constant $a$ increases) and a negative out-plane strain (distance between layers decreases), which leads to the redistribution of the electrons. As a consequence, the CBM, formed mainly by the hybridised Se $p_{x+y}$ orbitals and Pt $d_{xz+yz}$ orbitals, is lowered, while the VBM, formed mainly by the Se $p_z$ orbitals goes up, during which the band gap narrows. The compounds undergo semiconductor to metal transition with in-plane expansions with the same electronic origin as the layer number dependent bandgap.

\section{Acknowledgments}
Financial support from ``863''-project (2015AA034202) and Research Grant of Chinese Central Universities (ZYGX2013Z001) are acknowledged.

\end{document}